# Bright polariton OLEDs operating in the ultra-strong coupling regime

Armando Genco[1,2*], Alessandro Ridolfo[4], Salvatore Savasta[4], Salvatore Patanè[4], Giuseppe Gigli[1,3] and Marco Mazzeo[1,3*]

[1] Istituto di Nanotecnologia CNR-NANOTEC, Via Monteroni 73100, Lecce (Italy)

[2] Department of Physics and Astronomy, University of Sheffield, Sheffield S3 7RH, UK

[3] Dipartimento di Matematica e Fisica "Ennio De Giorgi", Università del Salento, Via per Arnesano, 73100 Lecce, Italy

[3] Dipartimento di Fisica e di Scienze della Terra, Università di Messina, Viale F. Stagno d'Alcontres 31, 98166 Messina - Italy

*Abstract:*

**The generation and control of exotic phenomena in organic electroluminescent microcavities, such as polariton lasing and non-linear optical effects, operating in strong and ultra-strong coupling regimes, is still a great challenge. The main obstacles originate from the small number of molecular classes investigated as well as from the absence of an efficient strategy aiming at the maximization of polariton states population. Here we report on bright polariton organic light emitting diodes made of a coumarin fluorescent dye emitting layer, working in the ultra-strong coupling regime up to a coupling strength of 33%. Owing to a high radiative decay emission, a large Stokes shift and a fine cavity-exciton tuning, the radiative pumping mechanism of polariton states has been fully optimized, leading a large portion (25%) of the emissive electrically pumped excitons to be converted in polariton emission. The resulting polariton OLEDs showed electro-optical performances up to 0.2% of external quantum efficiency and 700 cd/m$^2$ of luminance, corresponding to the highest values reported so far for this class of devices. Our work gives clear indications for an effective exploitation of organic polariton dynamics towards the development of novel quantum optoelectronic devices.**

*marco.mazzeo@unisalento.it

*armando.genco@unisalento.it

**Introduction**

The strong light-matter coupling (SC) regime in organic microcavities [1,2] allows to explore new physics and fascinating phenomena such as polariton lasing, [3] Bose-Einstein Condensation (BEC) [4,5] and superfluidity [6] even at room temperature. Furthermore, the ultra-strong coupling (USC) regime has been predicted to display intriguing and peculiar quantum electrodynamics features such as Casimir-like vacuum emission upon either non-adiabatic change of the coupling strength, [7] non-classical radiation from chaotic sources, [8] and new nonlinear optics effects. [9–11] Understanding how to increase brightness and efficiency of polariton light-emitting diodes (POLEDs) [12] working in SC and USC regime may led to the development of novel quantum optoelectronic devices and to reach electrically driven BEC at low thresholds. To this purpose, the maximization of the polariton flux (number of polaritons per unit area and per unit time) in the valley of the energy-momentum dispersion curve of the lower polariton states is a fundamental requirement. In organic disordered strongly-coupled microcavities both coherent polariton and incoherent exciton states coexist, the latter generally being the majority. [13] In this scenario, a non-resonant excitation of the system firstly gives rise to an initial population of the incoherent excited states (exciton reservoir). Then, the excitations are transferred to the coherent polariton states via two dominant processes (Figure 1a): i) radiative pumping of lower polariton (LP) states through photons emitted by the incoherent excitons via internal conversion and succeeding radiative decay channels; [13] ii) phonon scattering from the reservoir to the polariton states via non-radiative decay pathways assisted by the emission of vibrational quanta.[14] Compared to the inorganic counterpart, in the organic cavities an efficient phonon scattering is not easy to achieve, while radiative pumping has been demonstrated to be an efficient excitation transfer mechanism for optically-pumped organic polariton lasers.[3,15] A necessary condition for an efficient radiative pumping is a good overlap between the LP dispersion and the emission spectrum of the bare material. J-aggregates of organic dyes have been recently exploited to fabricate POLEDs, owing to their large electrical dipole moment, which allows to achieve SC regime despite the thin layer employed as active layer.[16] In that case the combination of a small Stokes shift and a high coupling ratio led to an overlap between the LP energy dispersion and the bare emission spectrum only at high angles, hindering an efficient radiative pumping. Indeed, the electroluminescence (EL) emission intensity in the SC regime was several times lower than the reference excitonic counterpart (OLEDs). On the other hand, our group recently demonstrated that polariton devices based on a thin emitting layer of a disordered squaraine dye in ultra-strong coupling regime [17–19] can reach the efficiencies of their weakly coupled reference devices.[20,21] Nevertheless, the quantum yield of the squaraine dye was poor, thus resulting in low absolute efficiencies in both the POLEDs and the reference device. An amorphous emitting layer of 2,7-bis[9,9-di(4-methylphenyl)-fluoren-2-yl]-9,9-di(4-methylphenyl) fluorene (TDAF) employed in metal cavity POLEDs was also recently reported in USC, showing an absolute maximum External Quantum Efficiency (EQE) of 0.1% and a maximum emitted optical power of about $3*10^{-4}$ mW/cm$^2$.[22] Despite the radiative pumping was favored by a partial LP spectral overlap with the uncoupled emission states, a ten-fold decrease of efficiency compared to similar standard OLEDs, reported elsewhere, was found.[23] In this work we report on a metal-cavity POLED architecture made by an amorphous layer of a coumarin fluorescent dye used as a novel strongly coupled emitter, namely *2,3,6,7-Tetrahydro-1,1,7,7,-tetramethyl-1H, 5H,11H-10-(2-benzothiazolyl) quinolizino-[9,9a,1gh] coumarin*, thereafter C545T. The large oscillator strength typical of this class of materials led to the occurrence of the USC regime with coupling strengths up to 33%. Moreover, the high brightness and the large Stokes shift of C545T allowed for an optimal polariton states population by radiative pumping. The results show that almost the 25% of emissive excitons generated by electrical pumping are converted in polaritons by their radiative decays, resulting in POLEDs with a maximum EQE of 0.2% and luminance of 700cd/m$^2$, that are, to the best of our knowledge, the highest values so far reported for this class of devices. As the coumarin molecules belong to a large class of materials successfully exploited for high performances OLEDs [24] and for solid-state laser dyes[25], we believe that our results may open the way to realize highly efficient polaritonic electroluminescent organic devices for broad photonic applications.

**Materials and Methods.**

We realized two classes of devices in USC regime, namely cavity-A and cavity-B with coupling strength factors G (the Rabi energy splitting normalized to exciton energy absorption) of 17% and 33%, respectively. Metallic and organic layers of our polariton devices have been thermally evaporated under high vacuum on glass substrates in a multi-chamber Cluster tool with the following order (**Figure 1b**): Al (25nm)/ p-doped layer (X nm)/ EBL (10 nm)/ C545T (20 or 55nm)/ HBL (10 nm)/ n-doped layer (X nm)/ Al (150nm). The *p*- doped layer has been realized by co-deposition of the dopant molecules of 2,3,5,6-tetrafluoro-7,7,8,8-tetracyanoquinodimetahne (F4TCNQ) into a matrix of N, N N', N'-tetrakis (4-methoxyphenyl)benzidine (MeOTPD); the *n* doped layer consists of 4,7-diphenyl,-1,10-phenanthroline (Bphen) doped with metallic Caesium (Cs). We varied the thickness (X) of both doped layers in order to tune the cavity length (detuning). EBL and HBL are the Electrons and Holes Blocking Layers, respectively. N,N′-Di-[(1-naphthyl)-N,N′-diphenyl]-1,1′-biphenyl)-4,4′-diamine (NPB) is the EBL, while the HBL is made by pure Bphen. The cavity devices have been compared to reference devices in which the bottom Al mirror electrode was replaced with an Indium Tin Oxide (ITO) anode of 150nm (Ref. device A and B). EL spectra were recorded using an Ocean Optics Charge-Coupled Device (CCD) spectrometer equipped with a Vis-NIR optical fibre (100 µm core size). The samples have been mounted on a rotating stage in order to perform the angular measurements. In order to measure the emitted power and to calculate the EQE, all the emission outcoupled from the glass substrate needs to be collected [26] by attaching the small area samples (15 mm$^2$) directly to a 1cm$^2$ Hamamatsu photodiode. The luminance was measured with a NIST calibrated Optronics OL770 spectrometer, coupled through an optical fibre, to a OL610 telescope unit.

**Results and Discussion**

We realized our devices by embedding a multi-layer p-i-n structure, comprising the emitting/coupling layer of C545T, between two metallic (Al) mirrors/electrodes (Fig. 1b).[20] In **Figure 1a** the absorption (Abs) and electroluminescence (EL) spectra of a thermally deposited C545T bulk film and of a reference ITO-OLED are respectively displayed. Abs and EL spectra feature two main peaks at 495 nm (2,51eV) and 571 nm (2,17eV), respectively. In Fig. 1a we show also the transmission spectrum of a 110 nm-thick cavity completely filled with the material, clearly displaying the upper and lower polariton peaks. The Rabi energy splitting was about 1.1eV with a remarkable coupling factor G value above 40%. The *p*- and *n*- doping of transporting layers reduce the ohmic losses to a negligible level, thus making charge injection almost independent from the nature of the electrodes employed, as well as from the thickness of the transport layers.[27]

As a first step for the optimization of radiative pumping, we fabricated strongly coupled electroluminescent microcavities with an active layer thickness of 55nm (cavity-A) which gives rise to a Rabi energy ($\hbar\Omega$, corresponding to half of the energy distance between UP and LP in Fig. 1a) of 430 meV, comparable with the molecular Stokes shift (∼ 370 meV). The resulting coupling ratio *G*=33% goes well beyond the threshold of USC regime.[17] A further analysis has been carried out by reducing the active layer thickness to 20nm (cavity-B samples) obtaining a Rabi energy of 220 meV, which corresponds to a coupling ratio *G*=17%. In this case, in order to maximize the radiative pumping, we have negatively detuned the cavities by increasing the thickness of the doped transport layers.

**Figure 2a** displays the EL intensities of LP emission as a function of the visual angle for both cavity-A and cavity-B at different detunings (about -100meV and -400meV respectively), corresponding to the best electro-optical performances (see inset in **Figure 3b** discussed later). As a first indication of the effectiveness of the radiative pumping process, we found that the angle-energy dependency of the EL emission intensity follows the energy distribution of reference OLEDs emission (shown on the right of Fig.2a), both regardless detuning and coupling strength. This is also clearly shown in more negatively detuned cavities (see Fig.S2). Figure 2a (as well as Fig.S2) also displays the LP dispersion curve (dashed white curves) calculated by solving the coupled harmonic oscillator Hamiltonian (see Supporting Information for details). In the figures, red and black dashed curves describe the resonance of the uncoupled cavity and the peak-position of the exciton distribution (obtained from the absorption of the bare molecular film), respectively. Compared to the cavity-A (G=33%), cavity-B (G=17%) produced high emission intensities in a wider angular range. The calculated Hopfield coefficients (see **Figure 2b**), which describe the photon ($|C|^2$) and exciton ($|X|^2$) fractions of the LP (see Supporting Information for details), suggest that this difference arises from a higher excitonic character of the LP states of cavity-A at high angles.

**Figure 2c** shows the EL peak intensities of cavity-B samples (orange square dots) plotted against the peak wavelength for different detunings (from -26meV to -729meV). These values were measured in the forward direction (at zero angle) at the same injected current (6.7mA/cm$^2$) and were normalized to the maximum emission of the reference OLEDs spectrum (the blue line curve in the figure refers to the reference OLED with an electroluminescent stack length of 140 nm). The variation of the doped transport layers thickness can slightly change the electroluminescence performances. In order to exclude from our analysis the effects of such electrical modifications on the POLEDs emission behaviour, we have normalized the EL intensities of Figure 2d, as well as each EQE of Figure 3b inset, to those of reference OLEDs. To furtherly demonstrate that the emission process is dominated by radiative pumping, we compared these data with those calculated by an optical simulation of the structures.[28–31] The model takes into account the light generation inside the optical structure, its partial reflections at interfaces, and the resulting interference effects.

In this theoretical framework, the organic emitting layer is modeled by a large number of mutually incoherent isotropic dipole antennas, whose emissions are weighted by the molecular EL spectral line-shape. The thicknesses and the optical constants of the employed materials have been evaluated by ellipsometric measurements (Fig. S1).[32] As in our model the phonon scattering is not considered, this theoretical approach is suitable to simulate polariton emission produced by only radiative pumping mechanism. As it can be clearly seen, the POLEDs peak intensities follow the reference EL emission spectrum profile and match very well the simulated data, confirming that the radiative pumping is the dominant emission mechanism. Little discrepancies between the computed and experimental data near 550nm, in both cavities and reference devices, can be attributed to some weak cavity effect at short wavelengths.

In **Figure 3a** we compared the current density and the luminance emitted in the forward direction by the optimized POLEDs. The lower current densities measured in cavities-A are attributed to an additional internal resistance caused by a thicker (un-doped) active layer used to increase the coupling strength. The maximum luminance of cavity-A and -B are about 550cd/m$^2$ (at 336mA/cm$^2$) and 700cd/m$^2$ (at 520mA/cm$^2$), respectively. It is worth noticing that this is the first time that luminance has been measured in polaritonic OLEDs, showing also remarkable values which are in line with the requirements for display or signalling applications.[33] Cavity-B showed a maximum current efficiency of 0.27 cd/A and a remarkable EQE of about 0.2% at a detuning of -445 meV, which are to our knowledge the best efficiencies reported so far for POLEDs. It is worth noticing that the maximum emitted optical power produced by C545T POLEDs (Fig.S3) is much higher than that showed by other molecules which showed BEC under optical pumping.[22] The increase of the coupling strength (cavity-A) results in a higher current efficiency (0.45 cd/A) because of a better overlap between LP spectrum and the uncoupled emission peak at zero-angle (Fig. 2a) and a lower current density flowing through the device. Nevertheless, as shown in Fig. 3b, for all the current densities its EQE is lower (0.13%), due to the EL suppression at higher angles (Fig. 2a). Despite the polariton EQEs reach record values for this class of devices, the reference OLEDs here shown have lower performances compared to typical standard OLEDs where the emitting dye was highly diluted in a host matrix. [24,34] Nevertheless, since the achievement of ultra strong coupling regime requires a large number of molecular dipoles coupled to the EM field ($\Omega_R \propto \sqrt{N_{dipoles}}$) and considering that there is an upper limitation in the thickness of the emitting layer in an OLED, the best compromise between light-matter coupling and electrical performances is achieved by depositing a thin and bulk layer of emitting dye.

In the inset of Fig. 3a the relative EQEs, calculated as the ratio between the EQE of each Polariton device with the related reference one, are plotted as a function of the cavity detuning. As cavity-B showed the best overlap for all the angles and a higher photonic Hopfield coefficient, it is not surprising that it shows the highest relative efficiency, emitting more than 25% of the out-coupled photons produced by the reference device. By dividing the number of photons per second per device area emitted in the forward direction to the LP photon fraction, [35] we calculated the polariton flux at the minimum energy of the LP branch (Fig. 3b). We found a comparable maximum polariton flux of $6*10^{11}$ cm$^{-2}$*s$^{-1}$ (at 336 mA/cm$^2$) and $5.85*10^{11}$ cm$^{-2}$*s$^{-1}$ (at 520 mA/cm$^2$) for cavity-A and -B, respectively. Nevertheless, at a fixed current density, cavity-A featured a higher polariton flux due to a lower LP photon fraction and to a better LP-emission spectral overlap at zero angle. Owing to the radiative pumping mechanism, where the number of generated polaritons is strictly related to that of photons emitted by weakly coupled exciton states, the efficiency of C545T POLEDs is directly proportional to those of the reference devices. This is opposite to POLEDs where phonon scattering is dominant compared to radiative pumping. In this case the portion of excitons that populate the polariton states does not depend on the whole number of injected excitons, resulting in a EQE independent of the reference OLED efficiency, as proved in our previous works employing a low efficiency squaraine molecule as active material. [21] Therefore, we argue that the phonon scattering play a minor role in this case.

**Conclusions**

In conclusion, we fabricated bright organic electroluminescent microcavities in the USC regime, consisting of a novel active layer of a C545T coumarin dye, optimized in order to increase the radiative pumping of strongly coupled states. By varying cavity detuning and coupling strength, it has been possible to achieve a good overlap between the bare emission spectrum and the lower polariton states, despite the high molecular Stokes shift, thus obtaining record efficiencies for this kind of devices. Owing to this process, a large portion of the emissive excitons generated by electrical pumping was converted in polariton emission (25% and ~ $6*10^{11}$ pol*$cm^{-2}$*$sec^{-1}$ in the minimum of LP branch). Specifically, the USC regime, together with the introduction of a slightly negative detuning, turns out to be beneficial for obtaining a high polariton populations at small angles. POLEDs with lower coupling strengths and lower detunings, but still in the USC regime, lead to higher EQE owing to a higher LP photon character in a broad range of angles. The high brightness and the high exciton-to-polariton radiative conversion efficiency shown by our polariton OLEDs give clear indications that coumarin dyes operating in USC regime can lead to a new class of stable organic photonic devices that can be exploited as a platform to study intriguing and exotic phenomena emerging at high light-matter interaction strengths.

**Acknowledgments**

We thank S. Carallo, G. Accorsi, V. Maiorano, A. Maggiore and M. Pugliese for technical and scientific support. This work was supported by Italian Ministry for Education, University and Research (MIUR) and European Union in the framework of 2007-2013 National Operational Programme for Research and Competitiveness (Smart Cities and Communities and Social Innovation call) under Grant PON04a3_00369, CUP: B35I12000180005.

# Figures

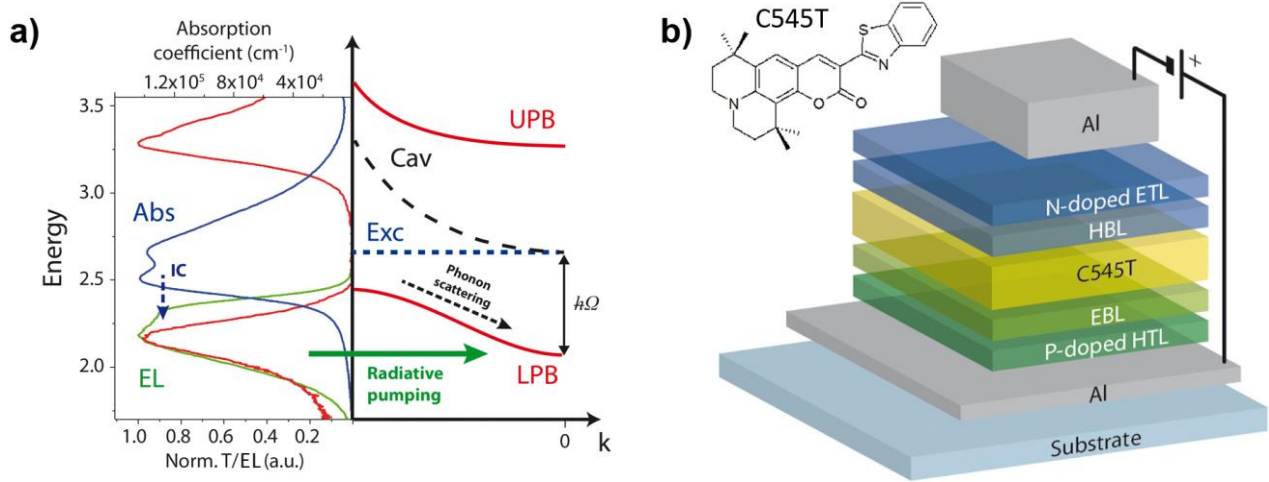

**Figure 1.** **(a)** On the left: absorption coefficient (blue line) of the C545T dye, normalized electroluminescence of a reference OLED (green line) and normalized transmittance (red line) of a microcavity in USC regime made of a 110nm thick film of C545T sandwiched between two thin Al mirrors (20nm); the arrows indicate the incoherent exciton emission pathways: they thermalize from absorption to emission states by Internal Conversion (IC, blue dashed arrow) and then decay radiatively emitting photons feeding the LPB states (radiative pumping, green arrow). On the right: schematic picture of the polaritons dynamics: UPB (upper polariton branch) and LPB (lower polariton branch) states are formed from the bare exciton (Exc) and cavity (Cav) modes. LPB can be populated by phonon scattering (black dashed arrow) as well as by radiative pumping; the energy difference between LP states and the bare cavity/exciton energy at the anticrossing, which for simplicity in this figure is considered at zero angle, is equal to $\hbar\Omega$. **(b)** Fabricated POLEDs layers scheme and chemical structure of C545T. Substrate is made of glass.

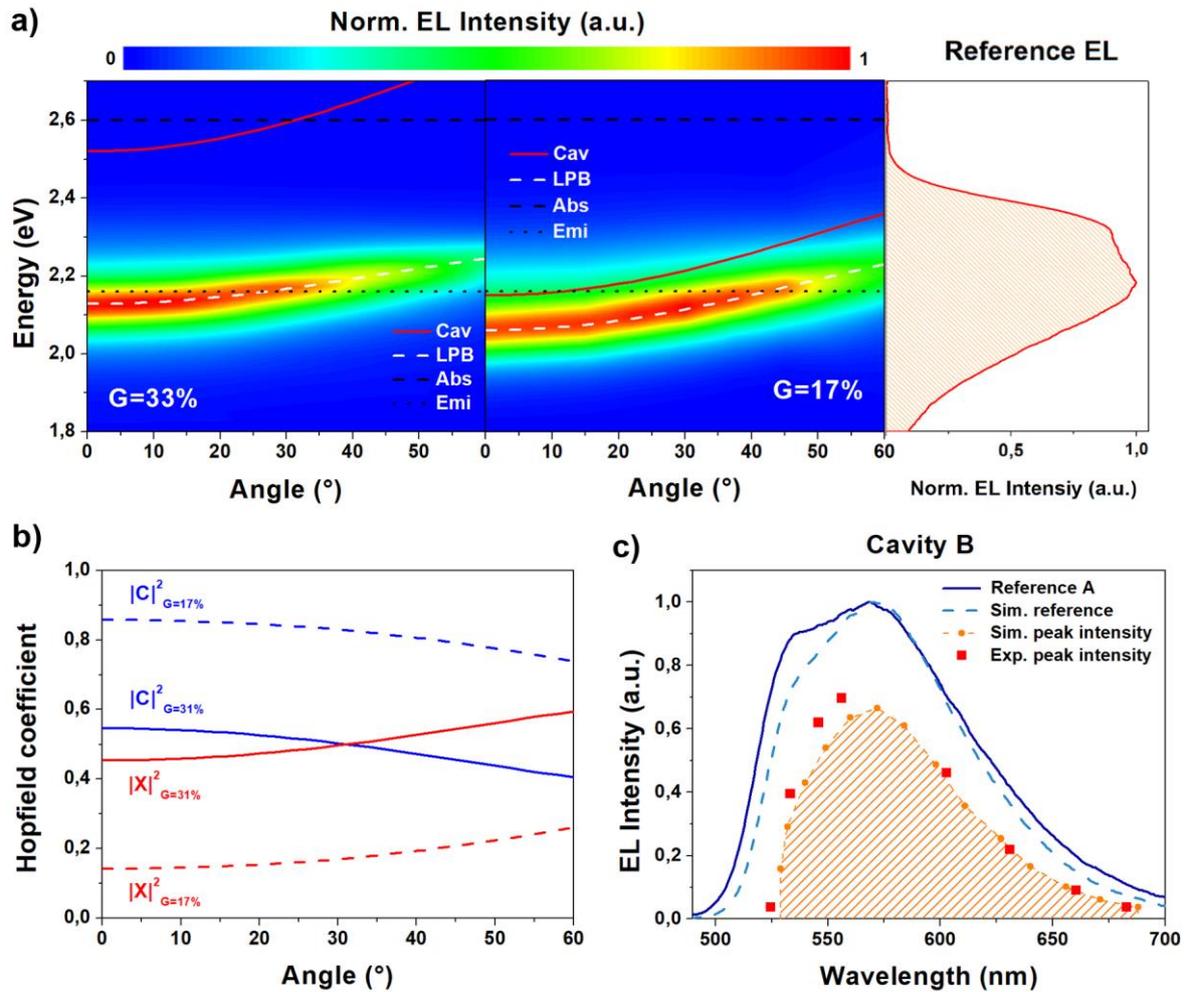

**Figure 2**. **(a)** EL intensity as a function of angle and energy of optimized POLEDs with *G*=33% and *G*=17% together (on the right) with the EL emission spectrum of a reference OLED at zero-angle. Dashed lines represent respectively the calculated bare cavity dispersions (red), the exciton absorption energy level (dotted black), the bare exciton emission energy level (dashed black) and the lower polariton branch (LPB, white). **(b)** Calculated photon (blue lines) and exciton (red lines) Hopfield coefficients for cavity-A (solid lines) and cavity-B (dashed lines). **(c)** Experimental and simulated EL spectra of reference OLEDs (blue solid line and cyan dashed line respectively); experimental and simulated EL peak intensities of cavity-B samples (red squares and orange circles respectively) at different detuning. The intensities are normalized to the references. All the measurements were taken at a fixed current (6.7mA/cm$^2$) and in the forward direction.

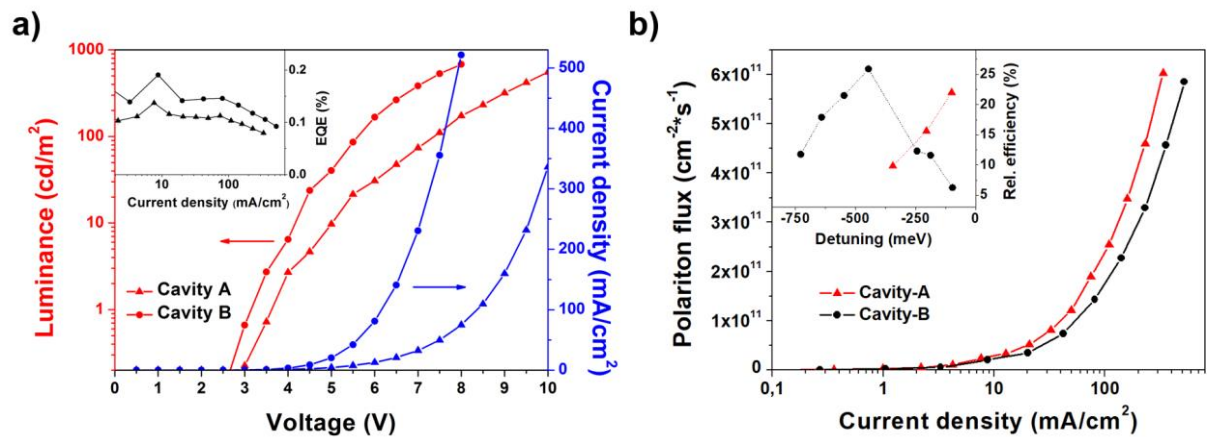

**Figure 3.** **(a)** Luminance and current density versus applied voltage of the two optimized POLEDs with different coupling strengths (cavity-A: triangles; cavity-B: circles). Inset: EQE of C545T POLEDs as a function of the current density. **(b)** Polariton flux against current density of the two optimized POLEDs (cavity-A: triangles; cavity-B: circles). Inset: relative efficiencies (normalized to reference OLEDs ones) of C545T POLEDs as a function of the cavity detuning.

# SUPPLEMENTARY MATERIAL

The calculated LP energies (white dashed lines in Fig.2a) are given by the following equation, obtained diagonalizing the Hamiltonian describing two coupled harmonic oscillators:

$$E_{LP} = \frac{E_0 + E_C(\theta)}{2} - \frac{1}{2}\sqrt{E_S^2 + (E_0 - E_C(\theta))^2} \quad , (1)$$

where $E_0$ is the exciton energy (black dashed line in Figs.2 a), $E_S$ is the Rabi energy splitting, $E_C(\theta) = E_C(0)/\sqrt{1 - sin^2\theta/n^2}$ is the angular dispersion energy of a bare cavity, $E_c(0)$ is the bare cavity energy at zero angle and n the fitted effective refractive index (n=2.1) of the medium embedded between the mirrors. The Hopfield coefficients were calculated following the equations reported by Deng et al. [1]:

$$|X|^2 = \frac{1}{2}\left(1 + \frac{(E_{exc} - E_{cav}(\theta))}{\sqrt{(E_{exc} - E_{cav}(\theta))^2 + 4(\hbar\Omega_R)^2}}\right);$$

$$|C|^2 = \frac{1}{2}\left(1 - \frac{(E_{exc} - E_{cav}(\theta))}{\sqrt{(E_{exc} - E_{cav}(\theta))^2 + 4(\hbar\Omega_R)^2}}\right),$$

where $E_{exc}$ is the exciton energy, $E_{cav}(\theta)$ is the cavity mode dispersion and $\Omega_R$ is the Rabi frequency.

The refractive index of the strongly coupled material considered in the optical model was measured by a J. A. Woollam spectroscopic ellipsometer. See in Figure S1 the optical constants calculated from the fit of the ellipsometric measurements. The dielectric constants of mirrors and of the other organic layers were taken from tabulated values [2].

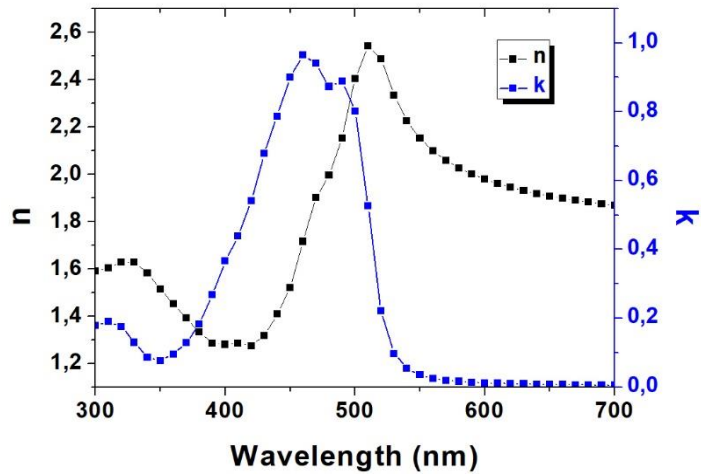

**Figure S1**: C545T optical constants calculated by fitting the ellipsometric measurements.

Figure S2(a,b) displays the EL intensities of Lower Polariton (LP) emission as a function of the angle for Cavity-A and Cavity-B samples at highly negative detunings. We observe that the maximum emission intensity always occurs at high angles near the emission peak of the C545T (black dotted lines) regardless the cavity coupling. This behaviour can be explained only assuming that the radiative pumping is the dominant emission pathway: polaritons are not able to scatter to the bottom of the LP branch faster than the standard decay routes of the incoherent states [3].

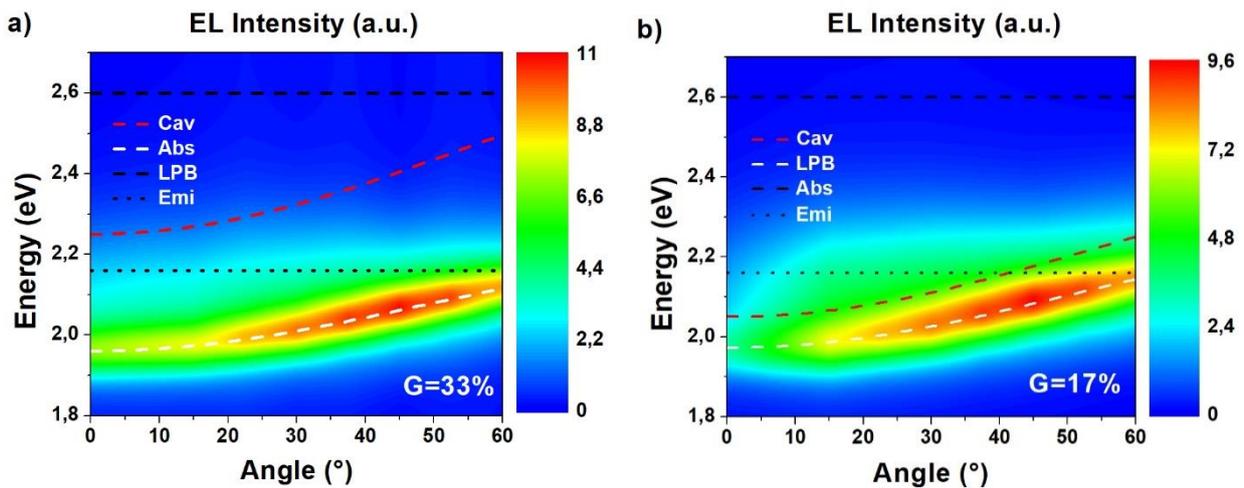

**Figure S2:** EL intensity as a function of angle and energy of POLEDs with G=33% (a) and G=17% (b) at a detuning of -350 meV and -555 meV respectively. Lines represent respectively the calculated bare cavity dispersions (red), the exciton absorption energy level (dashed black), the bare exciton emission energy level (dotted black) and the lower polariton branch (LPB, white).

Figure S3 shows the measured emitted optical power (mW/cm$^2$) as a function of the applied voltage for Cavity-A and Cavity-B samples.

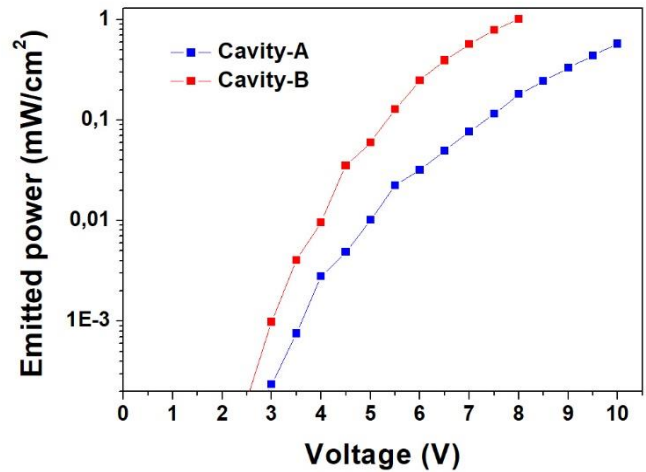

**Figure S3:** Emitted optical power, scaled by device area, as a function of increasing forward bias.

The figure below (Figure S4) shows the performances (luminance and current density against voltage) of the reference OLEDs with the same electroluminescent stack of the best cavities at the two different coupling strengths displayed in Figure3a.

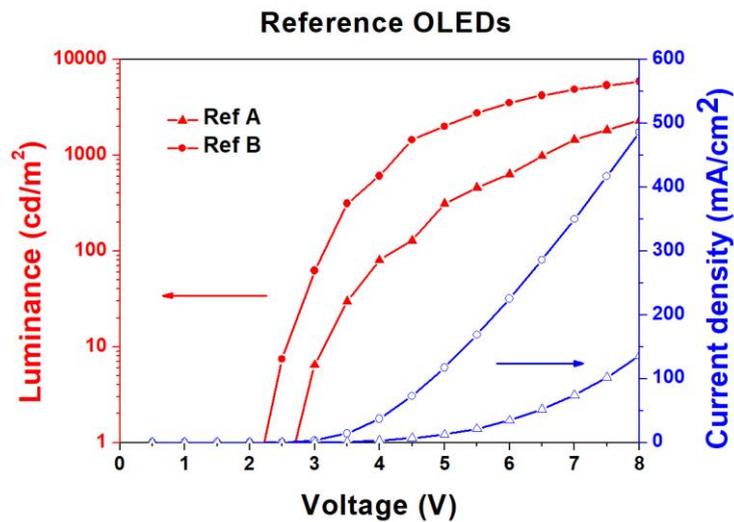

**Figure S4:** Luminance and current density versus applied voltage of the two reference OLEDs with the same EL stack of the optimized POLEDs with different coupling strengths (reference-A: triangles; reference-B: circles)

We also provide in Figure S5 an estimation of the transmission spectrum of each layer of the optimized cavity-A and cavity-B POLEDs calculated by transfer matrix simulations (using the experimental refractive index of each material measured by spectroscopic ellipsometry).

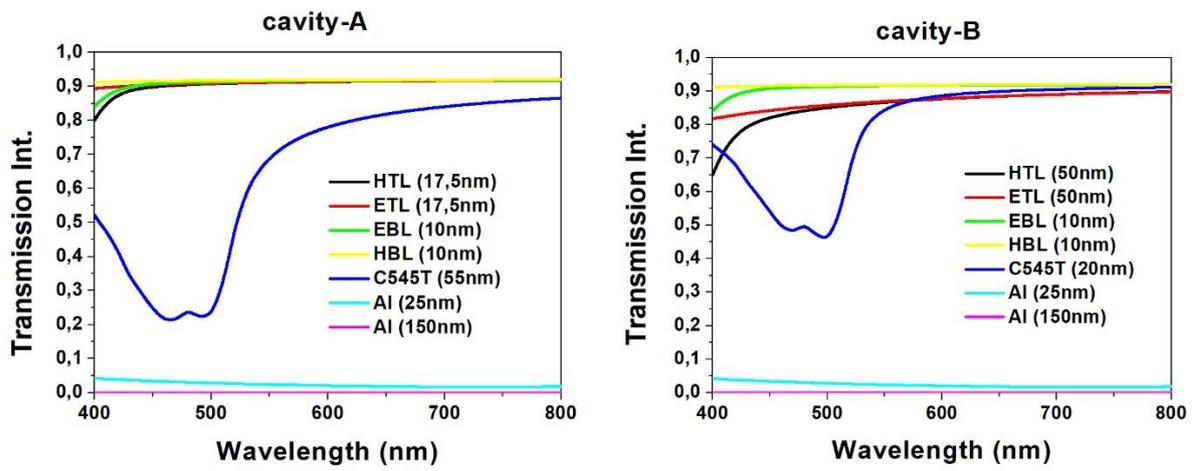

**Figure S5:** Transmission spectra of each layer of the optimized cavity-A and cavity-B POLEDs calculated by transfer matrix simulations.